\documentclass[%
onecolumn,
superscriptaddress,
amsmath,amssymb,
aps,
]{revtex4-2}

\usepackage{graphicx}
\usepackage{dcolumn}
\usepackage{bm}
\usepackage{hyperref}
\usepackage{verbatim} 
\hypersetup{
colorlinks,
citecolor=blue,
filecolor=blue, 
linkcolor=blue, 
urlcolor=blue
}

\usepackage{tikz}
\usetikzlibrary{arrows.meta, positioning, shapes.multipart, calc, fit}

\newcommand{\q}{{\mathbf q}}
\newcommand{\kk}{{\mathbf k}}
\newcommand{\G}{{\mathbf G}}

\renewcommand{\Im}{\mathrm{Im}} 
\renewcommand{\Re}{\mathrm{Re}}

\newcommand{\editor}[2]{%
  \expandafter\newcommand\csname #1note\endcsname[1]{%
    \textcolor{#2}{(\textbf{#1:} \textit{##1})}}%
  \expandafter\newcommand\csname #1\endcsname[1]{%
    \textcolor{#2}{##1}}%
  \expandafter\newcommand\csname #1cancel\endcsname[1]{%
    \textcolor{#2}{\sout{##1}}}%
\expandafter\newcommand\csname #1can\endcsname[1]{%
    \textcolor{#2}{\sout{##1}}}%
  \expandafter\newcommand\csname #1change\endcsname[2]{%
    \textcolor{#2}{\sout{##1} ##2}}%
      \expandafter\newcommand\csname #1ch\endcsname[2]{%
    \textcolor{#2}{\sout{##1} ##2}}%
  \newenvironment{#1text}{\color{#2}}{\color{black}}
}
\definecolor{Blu}{rgb}{0.00,0.00,1.00}
\definecolor{Red}{rgb}{1.00,0.00,0.00}
\definecolor{Cyan}{rgb}{0.00,0.50,0.50}
\definecolor{Green}{rgb}{0.00,0.70,0.00}

\editor{DAL}{Green}
\editor{CC}{Cyan}
\editor{kb}{orange}
\editor{KB}{orange}
\editor{R}{Red}

\renewcommand{\emph}{\textit}

\begin{document}


\title{Momentum- and frequency-resolved collective electronic excitations in solids: insights from spectroscopy and first-principles calculations}

\author{Dario A. Leon}
\email{dario.alejandro.leon.valido@nmbu.no}
\affiliation{
 Department of Mechanical Engineering and Technology Management, \\ Norwegian University of Life Sciences, NO-1432 Ås, Norway
}%

\author{Kristian Berland}
\affiliation{
 Department of Mechanical Engineering and Technology Management, \\ Norwegian University of Life Sciences, NO-1432 Ås, Norway
}


\begin{abstract}
Collective electronic excitations, including plasmons, excitons, and intra- and interband transitions, play a central role in determining the dynamic screening, optical response, and energy transport properties of materials. Recent advances in momentum- and frequency-resolved spectroscopies, such as electron energy-loss spectroscopy (EELS) and inelastic x-ray scattering (IXS), together with progress in first-principles many-body perturbation theory (MBPT) calculations, now allow collective
excitations to be mapped with considerable precision across the Brillouin zone. 
This topical review surveys current developments in the representation and interpretation of both experimental and theoretical dielectric-response spectra. Particular emphasis is placed on recent ways of representing 
spectral band structures (SBS) of the direct and inverse dielectric functions, such as analytical approaches based on multipole–Pad\'e approximants in momentum and frequency (MPA($\q$)), which provide a combined band-like description of the dispersion of the main collective excitations. 
We discuss how features observed in metals, semiconductors, and low dimensional systems reflect the interplay between electronic structure, screening strength, and local-field effects, and how post-processing procedures can improve the quantitative comparison between experiment and theory. Finally, we provide perspectives on open challenges and potential developments in quantitative dielectric-function analyses.
\end{abstract}

%
%

\maketitle

\section{Introduction}

Collective excitations such as plasmons, excitons, and intra- and inter-band transitions are fundamental quasiparticles in electronic materials. They originate from many-body interactions, reflecting how electrons dynamically screen each other in space and time.  
Understanding their dispersion, lifetimes, and coupling to the lattice or spin degrees of freedom is essential for interpreting spectroscopic data and engineering functional materials for photonics, plasmonics, and quantum technologies~\cite{Egerton2011book,Raether2006book,Haug-Koch2009book,Onida2002RMP,Mitrano2024PRX}.

These quasiparticles arise as solutions of the electronic response to external perturbations. 
Their properties are thus determined by  
the complex, momentum- and frequency-dependent dielectric response function $\varepsilon(\q,\omega)$ and its inverse $\varepsilon^{-1}(\q,\omega)$, which describe how the electronic system screens perturbations at a given wave vector $\q$ and frequency $\omega$. 
Peaks or poles in these response functions reflect the presence of self-sustained oscillations of the charge density, giving rise to intra- and inter-band transitions, plasmons, excitons, and mixed modes.

The frequency, broadening, and spectral weights of these poles are governed by the underlying electronic structure and screening properties of the many-electron system, and therefore manifest differently in metals, semiconductors, insulators, strongly correlated systems, and low-dimensional materials. 
The dielectric response provides a common theoretical framework for interpreting spectroscopic signatures of collective quasiparticles across materials with widely different electronic properties
~\cite{Fetter-Walecka1971book,Giuliani-Vignale2005book,Pines-Nozieres2005book,Onida2002RMP}.

While optical spectroscopies probe the long-wavelength limit, $\q \to 0$, many key aspects of screening and correlation emerge only at finite momentum transfer $\q$. The full $\varepsilon(\q,\omega)$ governs electron–electron and electron–hole interactions across the Brillouin zone, determining the dispersion of plasmons and excitons and the onset of intra- and inter-band continua, as well as the effective strength of local-field and exchange–correlation effects~\cite{Onida2002RMP,Giuliani-Vignale2005book,Mahan2013book,martin2016book}. 
Hence, moving from optical to momentum-resolved spectroscopies, such as electron energy-loss spectroscopy (EELS)~\cite{Egerton2011book} and inelastic X-ray scattering (IXS)~\cite{Burkel1991Book,Schulke2007book}, has opened a direct experimental window into the microscopic mechanisms of electronic screening.

Advances in both experimental instrumentation and theoretical modeling have substantially enhanced our ability to map and interpret these excitations. 
Monochromated and momentum-resolved EELS now routinely achieve sub-0.1~eV energy resolution and sub-nanometer probe sizes~\cite{Fink1989AEEP,Egerton2011book,Elgvin2025EELSReview}. 
In parallel, advances in synchrotron-based IXS instrumentation have significantly improved energy resolution, momentum stability, and signal-to-noise ratios, extending the reach of IXS to weak and high-energy electronic excitations with true bulk sensitivity
\cite{Baron2020Arxiv,Mitrano2024PRX}. 
Combined with advanced \emph{post-processing methods}, they enable the construction of experimental \emph{spectral band structures} (SBS), momentum–energy maps of the dielectric response along chosen $\q$-paths~\cite{Leon2024ZnO,Leon2025metals}. 

In parallel, first-principles methods based on the random-phase approximation (RPA), time-dependent density-functional theory (TD-DFT), and many-body perturbation theory (MBPT) approaches such as the Bethe–Salpeter equation (BSE) and beyond, offer accurate predictions of $\varepsilon(\q,\omega)$ and $\varepsilon^{-1}(\q,\omega)$. In turn, they enable quantitative comparisons of the many-body features that can be inferred from both theoretical and experimental spectra~\cite{martin2016book,Gatti-Sottile2013PRB,Alkauskas2013PRB,Fugallo2015PRB,Leon2024ZnO}. Recent studies on ZnO~\cite{Leon2024ZnO} and elemental metals~\cite{Leon2025metals} have demonstrated how such comparisons of SBSs can help uncover subtle many-body effects, including plasmon renormalization, excitonic effects, and mode mixing. 

As experimental and theoretical spectra become increasingly precise and accurate, not only effective post-processing methods, but also representation schemes become decisive for leveraging this information content to identify and characterize the underlying many-body features. 
The extraction of dielectric functions from experimental spectra typically involves several layers of transformations, including multiple-scattering deconvolution, background subtraction, normalization, and Kramers–Kronig inversion, each introducing their own uncertainties and interpretive biases~\cite{Egerton2011book,Leon2024ZnO,Elgvin2025EELSReview,Burkel1991Book,Schulke2007book}. 
Whether obtained from theory or experiment, SBSs are often highly complex, making concise analytical representations essential to move beyond qualitative inspection of numerical data and capture the underlying character of the collective quasiparticles. 

The paper is organized as follows:  
Section~\ref{sec:collective_qp} presents
the main collective excitations found in solids; Section~\ref{sec:fundamentals_response}
summarizes the theoretical fundamentals of the dielectric response; Section~\ref{sec:experiments} reviews experimental probes and data-reconstruction methods; Section~\ref{sec:representation} discusses spectral representations and momentum-resolved analysis; Section~\ref{sec:future} offers perspectives on open challenges and emerging directions; and Section~\ref{sec:conclusions} presents the conclusions.

\section{Collective quasiparticles in solids}
\label{sec:collective_qp}
The low-energy excitation spectrum of condensed matter is governed by a variety of 
\emph{collective quasiparticles}, emergent modes arising from the correlated dynamics of many electrons and nuclei~\cite{Hopfield1958PR,Pines1999excitations_solids,Giuliani-Vignale2005book}. 
These excitations determine the 
fundamental screening, optical, and transport properties of materials observed in momentum- and frequency-resolved spectroscopies, such as electron energy-loss spectroscopy (EELS), inelastic x-ray scattering (IXS), and different optical spectroscopies~\cite{Wooten1972book,Egerton2011book,Raether1988book,Giuliani-Vignale2005book,martin2016book}. 
 
\subsection{Plasmons}
Plasmons are collective oscillations of the electronic charge density, first identified in early theories of metallic screening and the interacting electron gas~\cite{Nozieres1958PR,Pines-Nozieres2005book,Fetter-Walecka1971book,Giuliani-Vignale2005book}.
Within linear-response theory, longitudinal plasmons are associated with resonances of the dielectric response or formally with poles of the inverse dielectric function.
In the weak-damping limit, these modes are often identified operationally by the approximate condition
\begin{equation}
\Re\!\left[\varepsilon(\q,\omega_{\mathrm{pl}}(\q))\right] \approx 0,
\label{eq:plasmon}
\end{equation}
which provides a convenient criterion for plasmon dispersions in both theory and experiment. 
When damping is strong or when coupling to interband continua is significant, plasmons are more accurately identified as broadened resonances in 
$\Im[-\varepsilon^{-1}(\q,\omega)]$. 
In the long-wavelength limit $\q \to 0$, this criterion in Eq.~\eqref{eq:plasmon} yields the familiar plasma frequency of a homogeneous electron gas~\cite{Pines-Nozieres2005book,Fetter-Walecka1971book,Giuliani-Vignale2005book},
\begin{equation}
\omega_{\mathrm{pl}} = \sqrt{\frac{4\pi \rho_e e^2}{m_e}},
\label{eq:plasma_frequency}
\end{equation}
where $\rho_e$ is the valence-electron density. Physically, this mode corresponds to an in-phase oscillation of the total electronic charge against the positively charged background.

At finite momentum $\q$, the plasmon acquires a dispersion reflecting the nonlocality of electronic screening.
Within the random-phase approximation (RPA) for a free-electron metal, the small-$\q$ behavior is quadratic~\cite{Pines-Nozieres2005book,Giuliani-Vignale2005book},
\begin{equation}
\omega_{\mathrm{pl}}(\mathbf{q}) \simeq
\omega_{\mathrm{pl}}
+ \frac{3}{10}\frac{v_F^2 q^2}{\omega_{\mathrm{pl}}}
+ \mathcal{O}(q^4),
\end{equation}
where $v_F$ is the Fermi velocity.
In real materials, band-structure effects, local-field corrections, and interband transitions substantially modify this result, with features such as damping, anisotropy, and mode hybridization. Experimentally, observed plasmons range from bulk and surface plasmons in simple metals to confined, anisotropic, or strongly damped modes in layered and low-dimensional systems~\cite{Raether2006book}. 


\subsection{Excitons}
Excitons are bound electron-hole pairs formed by the Coulomb attraction between an excited electron and the corresponding hole left in the valence band. They constitute neutral but strongly correlated two-particle excitations that dominate the optical response of semiconductors and insulators~\cite{Mahan2013book,martin2016book}. Depending on their spatial extent and binding energy, excitons are commonly classified as Wannier-Mott (delocalized, weakly bound) or Frenkel (localized, strongly bound)
types~\cite{Ueta2012book}. 
However, in many materials such as molecular crystals, layered compounds, and van der Waals heterostructures, excitons may exhibit significant charge-transfer or spatially indirect character, with partial real-space
or orbital separation of the electron and hole.  
These excitations interpolate between the Wannier-Mott and Frenkel limits and do not fit cleanly into either classification~\cite{Agranovich2008book,
Chernikov2014PRL,Wang2018RMP,Rivera2015NatComm,Fogler2014NatComm}.

In the Wannier-Mott regime, a hydrogenic model provides a useful estimate of the binding energy~\cite{Wannier1937PR,YuCardona2010book,Mahan2013book},
\begin{equation}
E_{\mathrm{b}} \simeq
\frac{\mu e^4}{2(4\pi\varepsilon_0)^2\hbar^2\,\varepsilon_\infty^2},
\end{equation}
where $\mu$ is the reduced electron-hole mass and $\varepsilon_\infty$ the electronic (high-frequency) dielectric constant. A quantitatively accurate description of excitons in real materials is obtained from the BSE equation~\cite{Rohlfing2000PRB,Onida2002RMP,martin2016book}, which determines exciton energies $E_\lambda$ and
amplitudes $A^\lambda$ via
\begin{equation}
\sum_{v'c'\kk'} 
\Big[
(E_{c\kk} - E_{v\kk})\,\delta_{vv'}\delta_{cc'}\delta_{\kk\kk'}
 + K^{eh}_{vc\kk, v'c'\kk'}
\Big]
A^\lambda_{v'c'\kk'}
= E_\lambda A^\lambda_{vc\kk} ,
\end{equation}
where the indices run over valence $v$, conduction $c$, and momentum $\kk$, and the electron-hole kernel $K^{eh}=v-W$ accounts for the bare, unscreened Coulomb exchange interaction $v$, and the screened Coulomb attraction $W$ between the excited electron and the hole. 
Excitons manifest as resonant pole contributions to the dielectric function,
\begin{equation}
\varepsilon(\omega)
= \varepsilon_\mathrm{b}(\omega)
+ \sum_{\lambda}
\frac{|S_\lambda|^2}{E_\lambda - \hbar\omega - i\gamma_\lambda},
\end{equation}
with energy $E_\lambda$, linewidth $\gamma_\lambda$, and oscillator strength $S_\lambda$.

At finite momentum $\q$, excitons acquire dispersion and can hybridize with plasmons or phonons. While this coupling is generally weak in the optical $\q \to 0$ limit, due to symmetry constraints and selection rules, it becomes increasingly relevant at finite $\q$. Hybridization can lead to the formation of mixed exciton-plasmon and exciton-phonon polaritons, altering the mode dispersion. 
Such effects are particularly pronounced in polar semiconductors~\cite{YuCardona2010book} and in reduced-dimensional systems, where weakened screening enhances electron-hole 
correlations~\cite{Agranovich2008book,Chernikov2014PRL,Wang2018RMP}. In polar semiconductors, long-range Fröhlich coupling to longitudinal optical phonons provides an additional channel for exciton dressing and damping, leading to phonon sidebands and temperature-dependent linewidths in
$\varepsilon(\q,\omega)$~\cite{YuCardona2010book,Adamska2021PRB}.

Fine-structure splittings, temperature-dependent linewidths, and phonon-assisted excitations are higher order effects that can be observed in $\varepsilon(\q,\omega)$.  
Theoretically, they can be captured by extending the BSE approach to include spinorial formulations with spin–orbit coupling and explicit exciton–phonon interactions~\cite{Marsili2021PRB,Adamska2021PRB,Mauri2025Arxiv}.


\subsection{Phonons and polaritons}
Phonons, the quantized normal modes of lattice vibrations, constitute a second major class of collective excitations~\cite{Baroni2001RMP,martin2016book}. 
Their coupling to electronic degrees of freedom underlies many-body phenomena such as polaron formation, superconductivity, and carrier relaxation~\cite{Giustino2017RMP,Devreese2009RPP,Carbotte1990RMP}. In insulating and semiconducting crystals, the interaction between lattice displacements and long-range electric fields is quantified by the Born effective charge tensor 
$Z^{*}$~\cite{BornHuang1954book,Gonze1997PRB}, defined through the induced macroscopic polarization
\begin{equation}
P_\alpha = \frac{e}{\Omega} \sum_{\kappa\beta} 
Z^{*}_{\kappa,\alpha\beta} \, u_{\kappa\beta},
\end{equation}
where $u_{\kappa\beta}$ is the displacement of atom $\kappa$ in the direction $\beta$ and $\Omega$, the cell volume.  
In polar materials, this coupling produces a longitudinal-transverse optical (LO-TO) splitting at $\q \to 0$ and gives rise to infrared-active \emph{phonon-polaritons}~\cite{Cochran1959PRL,Caldwell2015Nanophotonics}.

At higher energies, analogous coupling between electromagnetic fields and electronic collective modes leads to plasmon- and exciton-polaritons~\cite{Wooten1972book,book_Palik1985,YuCardona2010book,Hopfield1958PR}.
These mixed light-matter excitations emerge as resonances of the full electromagnetic response and share a common coupled-mode structure.

\subsection{Interplay and unified description}
Despite their distinct microscopic origins, plasmons, excitons, and phonons admit a common description within linear-response theory~\cite{Giuliani-Vignale2005book,martin2016book}.  
In the absence of local-field effects,
the longitudinal dielectric function reduces to a scalar quantity
\begin{equation}
\varepsilon(\q,\omega)
= 1 - v(\q)\,\chi(\q,\omega),
\end{equation}
which describes the coupled electronic and ionic response through the interacting polarizability $\chi$.
Mathematically, collective modes correspond to poles of the inverse dielectric function~\cite{Pines-Nozieres2005book,Giuliani-Vignale2005book}, or equivalently to zeros of $\varepsilon(\q,\omega)$ in the complex frequency plane,
\begin{equation}
\varepsilon(\q,\omega_\nu(\q)) = 0,
\end{equation}
which define the mode dispersions $\omega_\nu(\q)$.
Plasmons correspond to modes driven by electronic restoring forces~\cite{Giuliani-Vignale2005book}; excitons arise from resonant electron-hole attractions encoded in the BSE kernel
~\cite{martin2016book}; and phonons emerge from the ionic dynamical matrix modified by long-range Coulomb interactions through the Born effective charges~\cite{Baroni2001RMP}.  
Within this theory, all elementary excitations such as plasmons, excitons, and phonons appear as poles of the dielectric response, differing primarily in which degrees of freedom and interactions provide the dominant restoring force. 

Hybridization between different excitations arises when their spectral ranges overlap and finite coupling terms are present. A quantitatively reliable description of such mixed modes requires a microscopic theory that treats electronic, lattice, and electromagnetic degrees of freedom on equal footing.  TD-DFT or beyond approaches in MBPT offer suitable starting points~\cite{Baroni2001RMP,Onida2002RMP,martin2016book}. 
In practice, the full microscopic response is often highly structured and strongly frequency dependent.

Near isolated resonances, however, the resulting mixed modes can typically be captured by an effective coupled-oscillator description, which provides an intuitive low-energy representation of the hybridization effects without requiring an explicit microscopic treatment of the coupling
~\cite{Hopfield1958PR,Wooten1972book,YuCardona2010book}. 
\begin{equation}
\omega_{\pm}(\q) =
\tfrac{\omega_1(\q)+\omega_2(\q)}{2}
\pm
\sqrt{
\Big( \tfrac{\omega_1(\q)-\omega_2(\q)}{2} \Big)^2
+ (g(\q))^2
},
\end{equation}
where $\omega_1(\q)$ and $\omega_2(\q)$ are the uncoupled dispersions interacting through a coupling term $g(\q)$. This model leads to avoided crossings of the mixed modes $\omega_{\pm}(\q)$ and redistribution of spectral weights.

Exciton-plasmon coupling, phonon-polaritons, and other mixed modes thus emerge as different manifestations of the same underlying dielectric response. The common perspective provides the conceptual foundation for the momentum- and frequency-resolved representations discussed in the following sections, and motivates the use of SBSs and analytical models to disentangle complex experimental spectra.

\paragraph{Bridging experiment and theory.}
Advances in momentum-resolved EELS and IXS now allow direct access to the full $(\q,\omega)$ dependence of the dielectric response, enabling detailed mapping of mixed modes, damped continua, and correlation-driven collective excitations~\cite{Aryasetiawan1998RPP,Dagotto2005Science,Schulke2007book,Egerton2011book,Elgvin2025EELSReview}.  
Parallel developments in MBPT and TD-DFT provide quantitative predictions of response functions and their eigenmodes, narrowing the gap between experiment and first-principles theory~\cite{Aryasetiawan1998RPP,Reining2018wcms,Leon2024ZnO}. The subsequent sections build on this unified perspective 
to explain how excitations are measured, reconstructed, and represented in practice, with an emphasis on spectral band structures, analytical models, and materials-specific dielectric parametrizations.

\section{Fundamentals of the dielectric response}
\label{sec:fundamentals_response}
Understanding collective electronic excitations in solids requires a rigorous description of how electrons respond to time- and space-dependent external perturbations. This response is described by the \textit{dielectric function} $\varepsilon(\q,\omega)$, which connects the total and external electric fields through~\cite{Fetter-Walecka1971book,Giuliani-Vignale2005book}
\begin{equation}
\mathbf{E}_{\mathrm{tot}}(\q,\omega)
  = \frac{1}{\varepsilon(\q,\omega)}\,\mathbf{E}_{\mathrm{ext}}(\q,\omega).
\end{equation}
Within linear-response theory, $\varepsilon(\q,\omega)$ and its inverse $\varepsilon^{-1}(\q,\omega)$ are complex, frequency- and momentum-dependent functions whose zeros and poles reveal the presence of different \textit{collective modes} such as plasmons and excitons~\cite{Fetter-Walecka1971book, Giuliani-Vignale2005book, Mahan2013book, Pines-Nozieres2005book, martin2016book}.

\subsection{Microscopic and macroscopic dielectric functions}
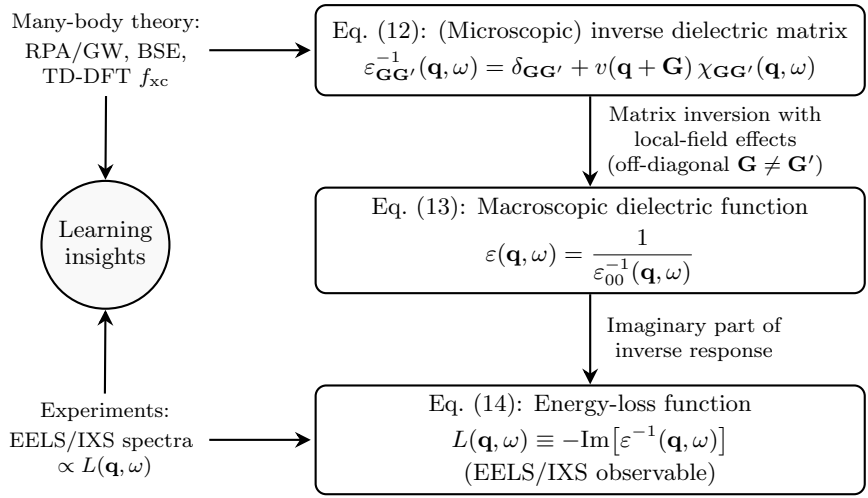
\begin{figure}[t]
  \centering
\begin{tikzpicture}[
  box/.style={draw, thick, rounded corners=4pt, minimum width=7.3cm, minimum height=1.2cm,
              align=center, font=\small, fill=white},
  arr/.style={-{Stealth[length=6pt,width=6pt]}, thick}, 
  darr/.style={<->, thick, >=Stealth}, 
  note/.style={font=\footnotesize, align=center}
]

  \node[box] (mic) {Eq.~(12): (Microscopic) inverse dielectric matrix\\[3pt]
    $\displaystyle \varepsilon_{\G\G'}^{-1}(\q,\omega)
    = \delta_{\G\G'} + v(\q+\G)\,\chi_{\G\G'}(\q,\omega)
    $};

  \node[box, below=1.2cm of mic] (macro) {Eq.~(13): Macroscopic dielectric function\\[3pt]
    $\displaystyle \varepsilon(\q,\omega)=
    \frac{1}{\varepsilon^{-1}_{0 0}(\q,\omega)}$};

  \node[box, below=1.2cm of macro] (loss) {Eq.~(14): Energy-loss function\\[3pt]
    $\displaystyle L(\q,\omega)\equiv -\Im\!\left[\varepsilon^{-1}(\q,\omega)\right]$ \\[2pt]
    (EELS/IXS observable)};

  \draw[arr] (mic) --
    node[right=0.1cm, note]{Matrix inversion with \\ local-field effects \\ (off-diagonal $\mathbf{G}\neq\mathbf{G}'$)}
    (macro);
  \draw[arr] (macro) --
    node[right=0.1cm, note]{Imaginary part of \\ inverse response} (loss);

  \coordinate (leftx) at ($(mic.west)+(-1.4cm,0)$); 

  \node[note, anchor=east] (mbnote) at ($(mic.west)+(-1.4cm,0)$) {Many-body theory:\\[2pt]
    RPA/GW, BSE, \\ TD-DFT $f_{\mathrm{xc}}$};
  \node[note, anchor=east] (expnote) at ($(loss.west)+(-1.4cm,0)$) {Experiments:\\[2pt] EELS/IXS spectra~{ } \\ $\propto L(\q,\omega)$};

  \draw[arr] (mbnote.east) -- ($(mic.west)+(-0.05,0)$);
  \draw[arr] (expnote.east) -- ($(loss.west)+(-0.05,0)$);

  \coordinate (midpt) at ($ (mbnote.south)!0.5!(expnote.north) $);
\node[
    draw,
    circle,
    thick,
    minimum size=1.4cm,
    font=\small,
    align=center,
    fill=gray!5
] (syn) at (midpt) {Learning\\insights};

  \draw[arr] (mbnote.south) -- (syn.north);
  \draw[arr] (expnote.north) -- (syn.south);

\end{tikzpicture}
  \caption{Relation between first-principles calculations at different levels of the many-body theory and the observables measured with EELS/IXS experiments. Conceptual flow linking the microscopic inverse dielectric matrix in Eq.~(12)\label{eq:12}, the macroscopic dielectric function in Eq.~(13)\label{eq:13}, and the energy-loss function in Eq.~(14)\label{eq:14}. In Eq.~(12), $\chi$ denotes the fully interacting (reducible) polarizability, and the dielectric matrix is restricted to the longitudinal response. Transverse electromagnetic effects are discussed separately in the context of polaritons.}
  \label{fig:dielectric_scheme}
\end{figure}
At the microscopic level, the dielectric response is described by the inverse dielectric matrix 
$\varepsilon^{-1}_{\mathbf{G}\mathbf{G}'}(\q,\omega)$, where $\mathbf{G}$ and $\mathbf{G}'$ are reciprocal lattice vectors and $\q$ is the momentum transfer within the first Brillouin zone. Local-field effects, arising from microscopic inhomogeneities in the crystal, are contained in the off-diagonal elements $\mathbf{G}\ne\mathbf{G}'$~\cite{Onida2002RMP,Giuliani-Vignale2005book,martin2016book}.
For longitudinal external fields, the \textit{macroscopic dielectric function} $\varepsilon(\q,\omega)$ is obtained by inverting the dielectric matrix and extracting its head element (the $\mathbf{G}=\mathbf{G}'=0$ component). Experimentally, the observable directly accessible in optical, EELS, and IXS spectroscopies is not $\varepsilon$ itself, but the imaginary part of its inverse through the \textit{energy-loss function} $L(\q,\omega)$ defined in Eq.~(14) of Fig.~\ref{fig:dielectric_scheme}.  
Figure \ref{fig:dielectric_scheme} summarizes the conceptual link between microscopic response functions computed from MBPT and the macroscopic quantities accessed in EELS and IXS.

\subsection
{Theoretical frameworks}
Most first-principles calculations of $\varepsilon(\q,\omega)$ start from the \textit{random-phase approximation}, RPA, in which the interacting response
$\chi$ is computed perturbatively from the independent-particle polarizability $\chi_0$ in the Dyson equation:
\setcounter{equation}{14}
\begin{equation}
\chi = \chi_0 + \chi_0 K \chi,
\end{equation}
where $\chi_0$ is usually computed from Kohn-Sham DFT states~\cite{Aryasetiawan1998RPP,Baroni2001RMP,Onida2002RMP,martin2016book}, and the interaction kernel is given by $K=v$ in RPA and includes exchange–correlation contributions beyond. RPA captures collective charge oscillations in simple metals with good accuracy, as exemplified in the comprehensive survey of elemental metals of Ref.~\cite{Leon2025metals}, but it neglects exchange-correlation and excitonic effects that are essential in semiconductors and insulators~\cite{Aryasetiawan1998RPP,Onida2002RMP,martin2016book}. In the ZnO study of Ref.~\cite{Leon2024ZnO}, comparison of RPA with hybrid-functional and BSE calculations shows that the inclusion of the electron-hole interaction redistributes the spectral weight near the absorption edge and the energy position of the plasmon, highlighting the role of the plasmon-exciton coupling.

Beyond RPA, TD-DFT introduces an exchange-correlation kernel $f_{\mathrm{xc}}(\q,\omega)$, and MBPT within the GW~+~BSE framework explicitly treats quasiparticle corrections and excitonic correlations~\cite{martin2016book}.
The extent to which these corrections shift the peaks in
$\varepsilon^{-1}(\q,\omega)$ or modify linewidths controls the balance between single-particle transitions and collective modes.

\subsection{Physical constraints}
The real and imaginary parts of the dielectric function encode complementary physical information.  
$\mathrm{Re}[\varepsilon(\q,\omega)]$ governs screening and identifies conditions for collective oscillations where $\Re[\varepsilon] \approx 0$, while $\Im[\varepsilon(\q,\omega)]$ governs absorption and damping effects~\cite{Giuliani-Vignale2005book,martin2016book}.
Together, they determine plasmon dispersions and lifetimes.  
The \emph{inverse} dielectric function $\varepsilon^{-1}(\q,\omega)$, on the other hand, describes the screening of the Coulomb interaction by the induced charge density.

The real and imaginary parts of both $\varepsilon(\q,\omega)$ and $\varepsilon^{-1}(\q,\omega)$ are not independent, but are related through the Kramers–Kronig relations, reflecting causality~\cite{Egerton2011book,martin2016book}. Additional constraints known as \textit{sum rules} ensure the internal consistency of static properties with respect to integration over dynamic quantities
~\cite{Pines-Nozieres2005book,Mahan2013book,Bradlyn2024PRB}. The \textit{$f$-sum rule}
\begin{equation}
\frac{2}{\pi} \int_0^{\infty} \! d\omega\, \omega\,
   \Im\!\left[-\varepsilon^{-1}(\q,\omega)\right]
   = \omega_\mathrm{pl}^2,
   \label{eq:f-sum rule}
\end{equation}
links the total spectral weight with the plasma frequency $\omega_\mathrm{pl}$ derived from the valence-electron density $\rho_e$ in Eq.~\eqref{eq:plasma_frequency}.

\paragraph{Summary.}
The dielectric function provides a unifying framework connecting microscopic electron dynamics with experimentally observed spectra. In the next section, we turn to the experimental realization of EELS and IXS measurements, and show how normalization of the recorded intensity and other post-processing techniques reconstruct the direct and inverse dielectric functions for quantitative comparison with first-principles theory.

\section{Experimental probes and data reconstruction}
\label{sec:experiments}
Momentum-resolved EELS and IXS are the two principal techniques for mapping collective electronic excitations over both momentum and energy. In the non-resonant regime, both probes measure quantities directly related to the imaginary part of the inverse macroscopic dielectric function, with a formal link established through the 
fluctuation-dissipation theorem~\cite{Fetter-Walecka1971book,Mahan2013book}.
Together with modern post-processing and normalization schemes, these methods enable quantitative access to the complex dielectric response $\varepsilon(\q,\omega)$ and its inverse $\varepsilon^{-1}(\q,\omega)$ across
the Brillouin zone~\cite{Egerton2011book,Schulke2007book}.

\subsection{Electron energy-loss spectroscopy (EELS)}

Momentum-resolved EELS, performed in transmission or scanning-transmission electron microscopes (TEM/STEM), provides a direct experimental route to probe collective excitations in $(\q,\omega)$ space~\cite{Egerton2011book}.
In a typical transmission geometry, the measured double differential cross section can be written as 
\begin{equation}
\frac{\mathrm{d}^2\sigma^\mathrm{EELS}}{\mathrm{d}\Omega\,\mathrm{d}\omega}
\propto
\frac{1}{q^2}\,L(\q,\omega),
\end{equation}
where $\q$ is determined by the electron scattering angle and incident beam energy, and $L(\q,\omega)=-\Im[\varepsilon^{-1}(\q,\omega)]$ is the energy-loss
function. The $1/q^2$ prefactor is due to the long-range Coulomb interaction and emphasizes small-momentum scattering, making EELS particularly sensitive to collective modes such as plasmons.

Modern monochromated STEM instruments routinely achieve sub-0.1~eV energy resolution and nanometer-scale spatial precision, enabling four-dimensional datasets that combine energy, momentum, and real-space information~\cite{Egerton2011book,Elgvin2025EELSReview}. These advances have transformed EELS into a central tool for constructing momentum-energy maps of the loss function that can be compared directly with first-principles calculations.

However, quantitative EELS analysis requires careful mitigation of experimental artifacts. Multiple inelastic scattering must be removed, for example, via Fourier-log or
statistical deconvolution, while the intense zero-loss peak demands robust background modeling~\cite{Egerton2011book}.
Accurate calibration of energy dispersion and momentum transfer is essential when tracing dispersions across Brillouin-zone paths, and finite angular acceptance must be accounted for to avoid artificial peak broadening.
Together, these factors define the effective experimental resolution in $(\q,\omega)$ space.

\subsection{Inelastic x-ray scattering (IXS)}
Inelastic x-ray scattering provides a powerful complement to EELS by probing the dynamic structure factor $S(\q,\omega)$ with bulk sensitivity and precise momentum definition at large $\q$. In the nonresonant regime relevant for electronic excitations, the IXS cross section can be expressed as
\begin{equation}
\frac{\mathrm{d}^2\sigma^\mathrm{IXS}}{\mathrm{d}\Omega\,\mathrm{d}\omega}
\propto
S(\q,\omega) \equiv q^2\,L(\q,\omega),
\end{equation}
highlighting the complementary momentum dependence of IXS and EELS.

Although the energy resolution of IXS (typically 30-100~meV) is lower than that of state-of-the-art monochromated EELS in the low-energy regime, IXS offers negligible multiple-scattering effects and access to an extended momentum range.
These capabilities enable the exploration of high-momentum plasmons, interband excitations, and deep electronic states, and make IXS particularly suitable for bulk-sensitive measurements~\cite{Burkel1991Book,Schulke2007book,Ishii2013JPSJ,Rossi2025xray}. Combined EELS-IXS studies can therefore provide a combined description of electronic excitations from the optical to the x-ray regime, disentangling
surface and bulk contributions and connecting local-field effects across different energy scales~\cite{Schulke2007book,Elgvin2025EELSReview}.

\subsection{Post-processing and normalization}
\label{sec:postprocessing}
Reconstruction of the dielectric response from EELS or IXS data involves several processing stages~\cite{Egerton2011book,Schulke2007book}. 
Raw spectra are first converted to single-scattering distributions by deconvolution, followed by subtraction of background and zero-loss contributions.
The resulting spectra must then be normalized to an absolute intensity scale, either by reference to the elastic peak or by enforcing integral sum rules, such as the $f$-sum rule in Eq.~\eqref{eq:f-sum rule}, as commonly applied IXS studies
\cite{Schulke1988PRB,Schulke1989PRB,Schulke1995PRB,Watanabe2006BCSJ,Weissker2010PRB}. Subsequent Kramers-Kronig analysis or model-based inversion yields the complex dielectric function $\varepsilon(\q,\omega)$.

A limitation of the normalization based on the $f$-sum rule is that the energy interval in which the spectra are measured needs to be sufficiently large to respect the rule, which is not always the case. A significant development in this context is a $\q$-dependent normalization scheme introduced in momentum-resolved EELS studies of ZnO~\cite{Leon2024ZnO}. The method relies on a partial integration of the $f$-sum rule in 
Eq.~\eqref{eq:f-sum rule}, with normalization performed relative to a reference spectrum. In contrast to the full integration of Eq.~\eqref{eq:f-sum rule}, a partial integration with a finite maximum frequency $\omega_\mathrm{m}$ introduces a $\q$ dependence:
\begin{equation}
S_{\omega_\mathrm{m}}(\q) = \frac{2}{\pi} \int_0^{\omega_\mathrm{m}}  d\omega \omega L(\q,\omega)
   \label{eq:partial f-sum rule}
\end{equation}
that can be used to normalize the recorded EELS intensity, $I_\mathrm{EELS}$, as
\begin{equation}
L_\mathrm{EELS}(\q,\omega) = I_\mathrm{EELS}(\q,\omega) \frac{S_\mathrm{ref}(\q)}{S_\mathrm{EELS}(\q)},
   \label{eq:normalization}
\end{equation}
where the reference $S_\mathrm{ref}(\q)$ can be obtained, e.g., from first-principles calculations, and both $S_\mathrm{ref}(\q)$ and $S_\mathrm{EELS}(\q)$ are computed with the same frequency interval given by the selected $\omega_\mathrm{m}$. 

\begin{figure}
    \centering
    \includegraphics[width=0.99\linewidth]{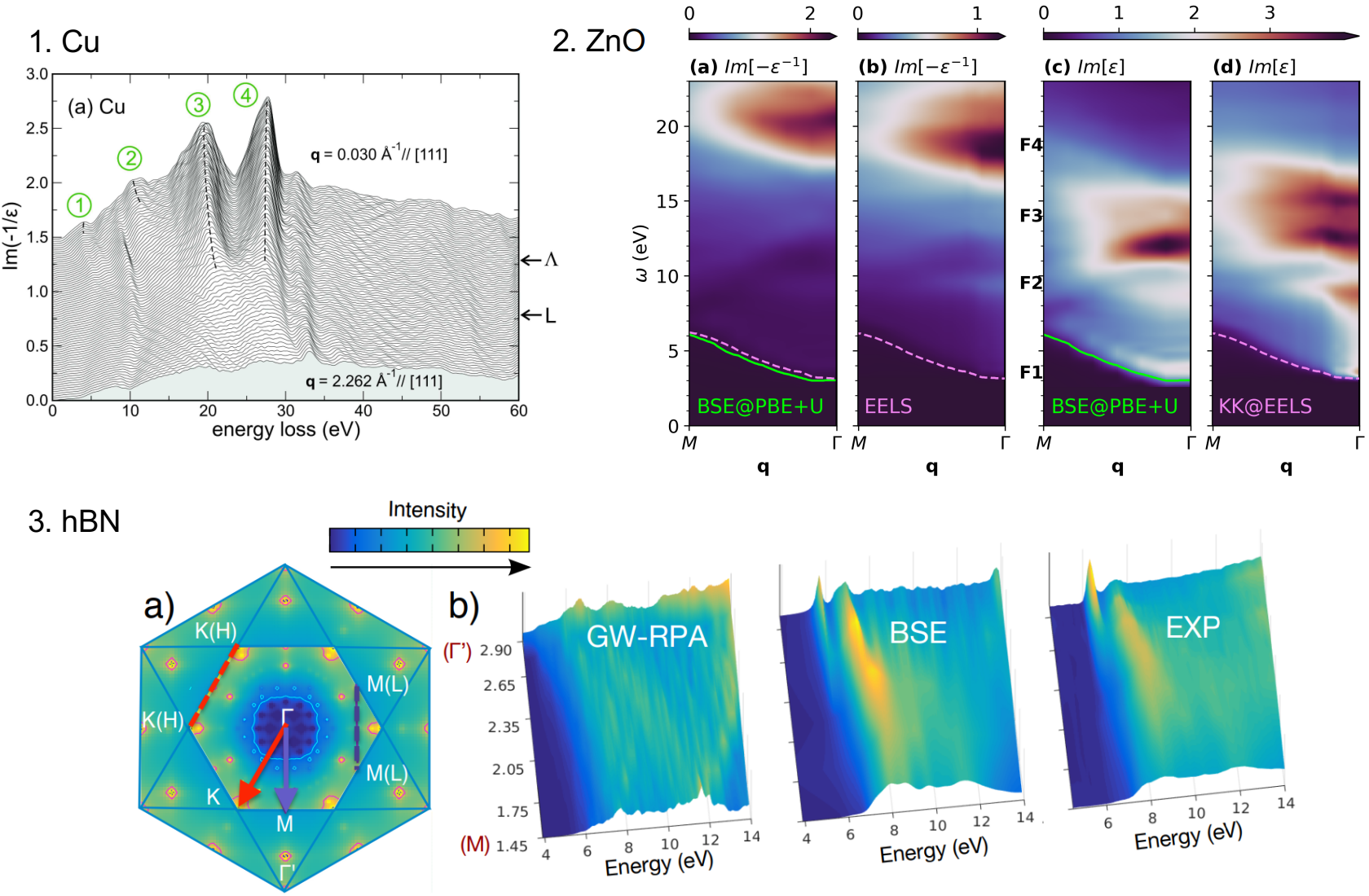}
    \caption{Examples of momentum and frequency dependent response functions of different materials. 1. Figure adapted from Ref.~\cite{Alkauskas2013PRB} (permission granted by license No. RNP/26/JAN/100835) corresponding to the loss function of bulk Cu obtained from REELS measurements.  2. Color-map representation of the bulk ZnO data from Ref.~\cite{Leon2024ZnO} (permission granted by license No. RNP/26/JAN/100838) corresponding to direct and inverse loss functions, $\Im[-\varepsilon^{-1}(\omega,\q)]$ and  $\Im[\varepsilon(\omega,\q)]$, computed at the BSE@PBE$+U$ level theort and compared to EELS measurements along the $M \Gamma$ direction. 3. Figure adapted from Ref.~\cite{Fugallo2015PRB} (permission granted by license No. RNP/26/JAN/100836) corresponding to the dynamical structural factor of hexagonal BN monolayer computed with GW and BSE and compared with IXS measurements. The main spectral features at low energies are indicated for Cu and ZnO.}
    \label{fig:collage}
\end{figure}

\subsection{Comparison with first-principles calculations}
Quantitative comparison between experiment and theory requires treating instrumental and post-processing effects consistently. Finite momentum and energy resolutions lead to peak broadening and intensity redistribution, while imperfect background subtraction or deconvolution can shift spectral weights and introduce artificial features. To ensure a consistent comparison, theoretical spectra are typically convolved with a broadening equivalent to that of the experiment. Moreover, the theoretical level must be sufficient to account for many-body interactions, such as quasiparticle corrections, local-field contributions, and excitonic effects, to reproduce the intrinsic line shapes. 
Quantitative dielectric-response calculations also require accurate equilibrium structures; for van der Waals bonded materials this typically necessitates a suitable choice of exchange-correlation functional, as interfragment separations can strongly affect band dispersions as well as plasmon and exciton energies~\cite{Rangel2016PRB,Li2025Vacuum}. 
Systematic modeling is necessary to distinguish between genuine physical effects encoded in the dielectric response and artifacts arising from experimental limitations or data processing, and thus provide reliable interpretation of the measured spectral features.

Figure ~\ref{fig:collage} illustrates representative cases in which first-principles calculations are confronted with momentum-resolved spectroscopy. These examples highlight how different materials classes: metals, semiconductors, and two-dimensional insulators, require distinct levels of theoretical treatment to achieve quantitative agreement with experiment. For bulk Cu~\cite{Alkauskas2013PRB}, the reconstructed loss function from reflection EELS (REELS) measurements captures both the free-electron plasmon and interband transitions once quasiparticle energies and local-field effects are included.
For ZnO~\cite{Leon2024ZnO}, comparison between experimental EELS and BSE-based calculations of the direct and inverse loss functions reveals the simultaneous presence of excitonic and plasmonic branches, demonstrating the need to account for electron–hole interactions in the screening.
In monolayer hBN~\cite{Fugallo2015PRB}, the momentum dependence of the dynamical structure factor measured by IXS is reproduced only after incorporating GW corrections and excitonic effects, underscoring the sensitivity of 2D materials to reduced screening.
Together, these case studies illustrate how controlled broadening, normalization, and many-body corrections enable quantitative and meaningful theory–experiment comparisons across a wide range of excitation energies and momentum transfers.

\paragraph{Best practices.}
For reproducible analysis and meaningful comparison between the experiment and $\varepsilon(\q,\omega)$ and $\varepsilon^{-1}(\q,\omega)$ obtained from first principles, we recommend: (i) making raw data on single-scattering spectra and instrumental response functions available, so that others can inspect and possibly even improve upon the processing method used; (ii) documenting deconvolution and background-removal parameters; and (iii) applying or reporting the normalization criteria employed for scaling of the spectra. Adoption of such practices will enhance reproducibility and enable a rigorous interpretation of SBS representations across experiments and materials.

\section{Representation and interpretation of spectral properties}
\label{sec:representation}
The dielectric function $\varepsilon(\q,\omega)$ and its inverse $\varepsilon^{-1}(\q,\omega)$ encode the full momentum- and frequency-dependent electronic response of a material~\cite{Onida2002RMP,Giuliani-Vignale2005book,martin2016book}. Extracting physical insights from these quantities, however, is not straightforward: the spectra obtained from EELS, IXS, or first-principles calculations often contain a rich superposition of intra- and interband transitions, plasmons, excitons, phonon-assisted excitations, and mixed modes. Interpreting such features becomes even more challenging when strong damping, local-field effects, or multiple excitation channels coexist, as is typical in correlated materials, polar semiconductors, and metals with valence electrons in localized orbitals, such as $d$ electrons. Effective representations are therefore essential to disentangle these contributions and extract qualitative and quantitative insights from the underlying dielectric response.

\subsection{Spectral band structures}
A practical approach for interpreting the dielectric response is to project the spectra along high-symmetry $\q$-paths, constructing \emph{spectral band structures}, SBSs, momentum–energy color maps of 
$\Im[\varepsilon(\q,\omega)]$ or $\Im[-\varepsilon^{-1}(\q,\omega)]$. These SBSs reveal the underlying landscape of electronic screening making the dispersion of elementary excitations explicit and providing a compact representation analogous to standard electronic band structures, but for bosonic response functions containing both collective modes and single-particle continua~\cite{Batson1983PRB,Schulke2007book,Egerton2011book,Leon2024ZnO,Leon2025metals}.

\begin{figure}
    \centering
    \includegraphics[width=0.99\linewidth]{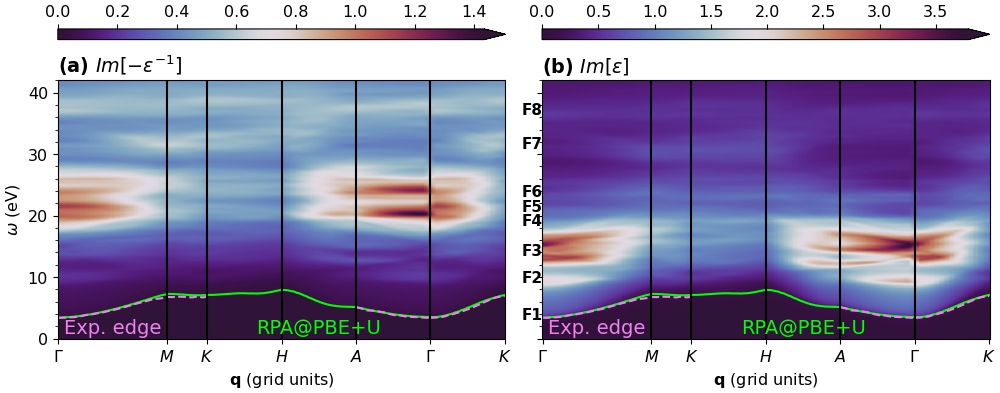}
    \caption{Color-map representation of the energy- and momentum-dependent $\Im[\varepsilon(\q,\omega)]$ (a) and its inverse $\Im[\varepsilon^{-1}(\q,\omega)]$ (b) along a high-symmetry $\q$-path in the Brillouin zone, replotted from the RPA data of Ref.~\cite{Leon2024ZnO} (permission granted by license No. RNP/26/JAN/100838). Experimental absorption edges are included where available.}
    \label{fig:ZnO}
\end{figure}

Figure~\ref{fig:ZnO} illustrates this approach for ZnO, showing SBSs of both $\Im[\varepsilon]$ and $\Im[\varepsilon^{-1}]$ computed within RPA and compared to available experimental onsets. The onset in the valence–conduction gap region and other single-particle, plasmonic, and mixed resonances, are all visible, highlighting the ability of SBSs to reveal momentum-dependent screening and the emergence of distinct excitation channels due to anisotropy.

\subsection{Analytical models of the dielectric function}
To complement the direct visualization of SBS maps, analytical models offer compact functional forms that parameterize the essential excitation channels. Drude–Lorentz oscillators~\cite{Wooten1972book} and model variations~\cite{Vos2017JPCS}, plasmon-pole approximations~\cite{Hybertsen1986PRB,vonderLinden1988PRB,Zhang1989PRB,Godby1989PRL,Engel1993PRB}, hydrodynamic models~\cite{Ritchie1957PR,Ciraci2012Science}, and other semiclassical approaches provide transparent connections between resonances, oscillator strengths, and screening mechanisms. Yet these models have limited generality, especially when multiple interband channels or strong momentum dependence are present.

A growing need therefore exists for representations that are (i) compact, (ii) physically interpretable, and (iii) quantitatively faithful to first-principles spectra across broad frequency and momentum ranges.

\subsection{Effective {\it ab initio} representations: the MPA($\q$) framework}
An emerging approach that satisfies these criteria is the
\emph{multipole–Padé approximant} (MPA) representation, which expresses the dielectric function and its inverse as rational functions in frequency space. Rational-function representations of response functions have a long
history in many-body physics, notably through Padé approximants used for analytic continuation \cite{Vidberg1977JLTP}. In the MPA representation, poles and residues encode the dominant collective and single-particle excitations. MPA was initially developed as an efficient approach to accelerate the performance of first-principle MBPT calculations~\cite{Leon2021PRB,Leon2023PRB,Leon2025PRB}. When constructed along a momentum path, the resulting MPA($\q$) yields a compact, band-structure-like description of the dielectric response.

\begin{figure*}[hbt!]
    \centering
    \includegraphics[width=0.266\textwidth]{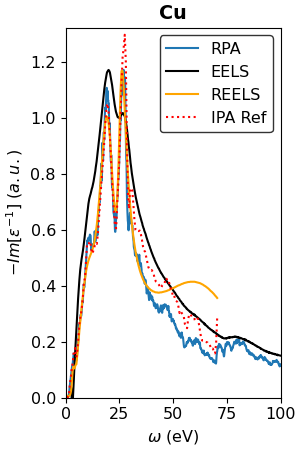}
    \includegraphics[width=0.2\textwidth]{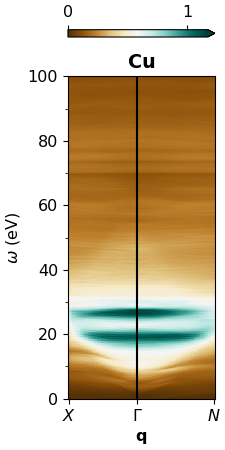}
    \includegraphics[width=0.465\textwidth]{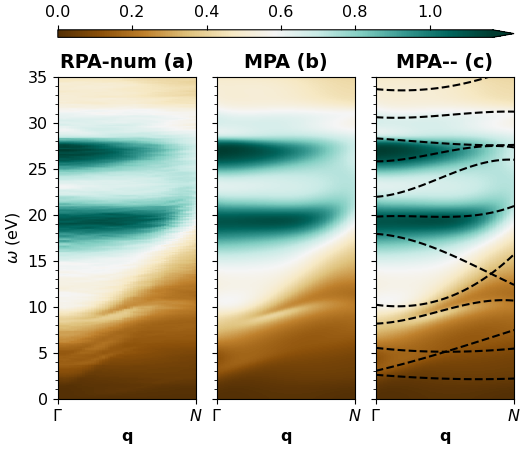}
    \caption{(left panel) Comparison of the computed RPA loss function of Cu in the optical limit with EELS, REELS measurements, and reference IPA results. The data is taken from Ref.~\cite{Leon2025metals}. (middle panel) Spectral $Y (\omega, \q)$ band structure of Cu computed from the numerical RPA results along the $X \Gamma N$ $\q$-path. (right panels) Spectral $Y (\q, \omega)$ band structure of Cu in the $\Gamma N$ direction reconstructed with MPA($\q$) with a number of $n_Y = 12$ poles.}
    \label{fig:Cu}
\end{figure*}

Figure~\ref{fig:Cu} demonstrates the effectiveness of the MPA($\q$) approach for Cu. The left panel compares the optical-limit loss function from RPA with EELS, REELS, and IPA references, while the center panel shows the full SBS
$Y(\omega,\q)$ along the $X\Gamma N$ path. The right panel displays the corresponding MPA($\q$) reconstruction using $n_Y = 12$ poles, which reveals a small number of well-defined multipole branches that capture the underlying plasmonic and intra- and interband structure with quantitative accuracy. The resulting representation condenses the complexity of the numerical spectra into an interpretable set of \emph{multipole bands}, enabling mode tracking, linewidth analysis, and systematic comparison across materials.

\paragraph{Summary.}
The combination of SBS representations and analytical reconstructions such as MPA($\q$) provides a powerful, general framework for understanding the momentum-dependent dielectric response across metals, semiconductors,
insulators, and low-dimensional systems. These tools reduce the complexity of $\varepsilon(\q,\omega)$ to a set of interpretable descriptors, enabling mode classification, quantitative comparison to experiment, and systematic
exploration of collective excitations in complex materials.

\section{Open challenges and future directions}
\label{sec:future}
The developments reviewed in this article, from SBS representations and multipole–Padé models to improved experiment–theory comparison protocols, highlight both the opportunities and the remaining challenges in achieving a fully quantitative understanding of momentum- and frequency-resolved excitations in solids. Below we outline several directions in which progress is particularly timely.

\begin{itemize}
    \item \textbf{Improved experimental post-processing.}  
    As emphasized in Sec.~\ref{sec:postprocessing}, the reconstruction of $\varepsilon(\q,\omega)$ and $\varepsilon^{-1}(\q,\omega)$ depends critically on multi-step procedures (background removal, 
    multiple-scattering deconvolution, normalization). Standardized analysis pipelines, publication of raw data and metadata, and community-agreed reference standards are essential to ensure consistent comparison across instruments, laboratories, and materials.
    
    \item \textbf{Quantitative linewidths and damping.}  
    The SBS and MPA($\q$) frameworks provide structured views of dispersive excitations, but extracting intrinsic lifetimes remains hindered by instrumental broadening, overlapping continua, and many-body scattering channels. More reliable deconvolution schemes, separation of intrinsic and extrinsic broadening mechanisms, and robust uncertainty quantification are needed to turn spectral maps into accurate lifetime benchmarks.

    \item \textbf{Beyond-RPA and finite-temperature theory.} 
    The applicability of SBS and MPA($\q$) analyses can be further expanded by incorporating dynamical exchange–correlation kernels, vertex corrections, and electron–phonon coupling directly into $\varepsilon(\q,\omega)$.  
    These ingredients are increasingly required to interpret experimental spectra that already exhibit temperature-dependent renormalization, mode hybridization, and strong damping beyond RPA predictions.

    \item \textbf{Data-driven interpretation and automation.}  
    The structured nature of SBS and MPA($\q$) representations makes them particularly amenable to machine-learning approaches. Potential applications include automated mode classification (plasmon/exciton/phonon-polariton), noise-resistant feature extraction, and fast surrogate models for many-body spectra. Such tools could streamline real-time analysis during EELS or IXS experiments. Such approaches will be most effective when combined with physical constraints, sum rules, and causality conditions that preserve the interpretability of the extracted modes.

    \item \textbf{Next-generation momentum-resolved spectroscopies.}  
    Advances in monochromated EELS, nano-focused IXS, and emerging ultrafast modalities will provide higher resolution and richer contrast. These capabilities will enable momentum-resolved analyses at smaller length scales, across interfaces and heterostructures, in operando environments, and on systems with strong anisotropy or disorder. Theoretical and representational tools, particularly analytical models such as MPA($\q$), must evolve to meet the demands of these new regimes.
    
    \item \textbf{Identification of emergent high-order modes.} 
    Momentum-resolved spectroscopies increasingly reveal hybrid excitations such as plasmon-exciton, plasmon-phonon, and polaronic modes, whose microscopic nature and classification remain open questions. Disentangling these excitations and establishing unambiguous physical criteria for their identification will be essential for advancing our understanding of screening and correlations in complex materials.
\end{itemize}

Overall, the combined use of momentum-resolved spectroscopies, rigorous dielectric-function reconstruction, and effective spectral representations offers a pathway toward a predictive, material-specific understanding of 
collective excitations. Continued development in each of the areas above will be essential for turning these methodologies into standard tools for exploring screening, correlations, and quasiparticle dynamics across modern condensed-matter systems.

\section{Conclusions}
\label{sec:conclusions}
In this topical review we have surveyed recent progress in the analysis of momentum- and frequency-resolved collective excitations in solids, with an emphasis on how \emph{representations} of the dielectric function shape the physical interpretation and classification of spectroscopic data. Spectral band structures, momentum-projected analyses, and analytical models such as multipole–Padé approximants offer complementary ways of organizing and visualizing the information contained in $\varepsilon(\q,\omega)$ and 
$\varepsilon^{-1}(\q,\omega)$, enabling clearer identification of plasmons, excitons, interband continua, and mixed modes across materials.

A central outcome of this review is the emergence of a unified methodological framework linking experiment and first-principles theory through consistent post-processing, normalization, and forward modeling. By treating instrumental resolution, background removal, and deconvolution on equal footing in both measured and computed spectra, meaningful quantitative comparisons become possible, revealing subtle many-body effects that would otherwise remain hidden.

Looking ahead, continued advances in high-resolution EELS and IXS, together with the systematic inclusion of excitonic, phononic, and temperature-dependent interactions in first-principles calculations, will further expand the scope of momentum-resolved dielectric-function analysis.
Combined with data-driven tools and increasingly robust spectral representations, these developments point toward a predictive, material-specific understanding of collective excitations and electronic screening in complex condensed-matter systems.


\section*{Funding}
This work is supported by the Research Council of Norway in the MORTY project (664350).

\section*{Roles}
D.A.L. conceptualized the paper, performed the initial literature survey, and wrote the first draft. Both D.A.L. and K.B. contributed to manuscript writing, literature survey, and editing.



\bibliography{biblio}

@article{Leon2024ZnO,
  title = {Unraveling many-body effects in ZnO: Combined study using momentum-resolved electron energy-loss spectroscopy and first-principles calculations},
  author = {Leon, Dario A. and Elgvin, Cana and Nguyen, Phuong Dan and Prytz, \O{}ystein and Hage, Fredrik S. and Berland, Kristian},
  journal = {Phys. Rev. B},
  volume = {109},
  issue = {11},
  pages = {115153},
  numpages = {17},
  year = {2024},
  month = {Mar},
  publisher = {American Physical Society},
  doi = {10.1103/PhysRevB.109.115153},
  url = {https://link.aps.org/doi/10.1103/PhysRevB.109.115153}
}

@misc{Leon2025metals,
   author = {Leon, Dario A. and Cardoso, Claudia and Berland, Kristian},
   title = {Bulk plasmons in elemental metals},
   howpublished = {\url{https://arxiv.org/pdf/2510.07261}},
   year = {2025}
}

@misc{Baron2020Arxiv,
   author = {Alfred Q. R. Baron},
   title = {Introduction to High-Resolution Inelastic X-Ray Scattering},
   howpublished = {\url{https://arxiv.org/pdf/1504.01098v7}},
   year = {2020}
}

@misc{Elgvin2025EELSReview,
  title={Advances in momentum-resolved EELS of phonons, excitons and plasmons in 2D materials and their heterostructures},
  author={Cana Elgvin and Fredrik S. Hage and Khairi F. Elyas and Katja Höflich and Øystein Prytz and Christoph T. Koch and Hannah C. Nerl},
  howpublished = {\url{https://arxiv.org/pdf/2510.07261}},
   year = {2025}
}

@misc{Mauri2025Arxiv,
   author = {Zachary N. Mauri and Christopher J. Ciccarino and Jonah B. Haber and Diana Y. Qiu and Felipe H. da Jornada},
   title = {Exciton-polaritons and exciton localization from a first-principles interacting Green’s function
formalism},
   howpublished = {\url{https://arxiv.org/pdf/2503.13613}},
   year = {2025}
}

@book{Egerton2011book,
  title={Electron Energy-Loss Spectroscopy in the Electron Microscope},
  author={Egerton, RF},
  year={2011},
  publisher={Springer},
  doi={10.1007/978-1-4419-9583-4}
}

@book{Schulke2007book,
    author = {Schülke, Winfried},
    title = {Electron Dynamics by Inelastic X-ray Scattering},
    publisher = {Oxford University Press},
    year = {2007},
    month = {06},
    isbn = {9780198510178},
    doi = {10.1093/oso/9780198510178.001.0001},
    url = {https://doi.org/10.1093/oso/9780198510178.001.0001},
}

@book{Burkel1991Book,
  author = {Burkel, E.},
  title = {Inelastic Scattering of X-Rays with Very High Energy Resolution},
  publisher = {Springer Berlin, Heidelberg},
  isbn = {1615-0430},
  doi = {10.1007/BFb0045861},
  year = {1991}
}

@book{Fetter-Walecka1971book,
  author={Alexander L. Fetter and John Dirk Walecka},
  title={Quantum theory of many-particle systems},
  publisher={McGraw-Hill, New York},
  year=1971,
}

@book{Wooten1972book,
        author = {Frederick Wooten},
        title = {Optical properties of solids},
        publisher = {Academic Press},
        year = {1972},
}

@inbook{book_Palik1985,
   title =     {Handbook of Optical Constants of Solids},
   author =    {Edward Palik},
   publisher = {Academic Press.},
   address =   {Institute for Physical Science and Technology 
University of Maryland College Park, Maryland },
   isbn =      {9780080556307},
   year =      {1985},
   series =    {},
   edition =   {1st ed},
   volume =    {1-4},
   pages =     {283},
}

@book{Pines-Nozieres2005book,
        author = {David Pines and Philippe Nozières},
        title = {The Theory of Quantum Liquids, Vol. 1: Normal Fermi Liquids},
        publisher = {Avalon Publishing},
        year = {1966},
}

@book{Mahan2013book,
        author = {Gerald D. Mahan},
        title = {Many-Particle Physics},
        publisher = {Springer New York, NY},
        year = {2013},
        doi = {10.1007/978-1-4757-5714-9}
}

@book{Raether1988book,
        author = {Heinz Raether},
        title = {Surface Plasmons on Smooth and Rough Surfaces and on Gratings},
        publisher = {Springer Berlin, Heidelberg},
        year = {1988},
        doi = {10.1007/BFb0048317}
}

@book{Raether2006book,
  title={Excitation of plasmons and interband transitions by electrons},
  author={Raether, Heinz},
  volume={88},
  year={2006},
  publisher={Springer}
}

@book{Giuliani-Vignale2005book,
        author = {Giuliani, Gabriele and Vignale, Giovanni},
        title = {Quantum Theory of the Electron Liquid},
        publisher = {Cambridge University Press},
        year = {2005},
        doi = {10.1017/CBO9780511619915},
}

@book{martin2016book,
  author={Martin,Richard M. and Reining,Lucia and Ceperley,David M.},
  title={Interacting Electrons},
  year=2016,
  publisher={Cambridge University Press},
  address={Cambridge}
}

@book{Pines1999excitations_solids,
author="Pines, David",
title="Elementary excitations in solids",
publisher="Perseus Books",
year="1999",
series="Advanced book classics",
doi="https://doi.org/10.1201/9780429500855"
}

@book{Haug-Koch2009book,
author = {Haug, Hartmut and Koch, Stephan W},
title = {Quantum Theory of the Optical and Electronic Properties of Semiconductors},
publisher = {WORLD SCIENTIFIC},
year = {2009},
doi = {10.1142/7184},
address = {},
edition   = {5th},
URL = {https://www.worldscientific.com/doi/abs/10.1142/7184}
}

@book{Ueta2012book,
author = {Masayasu Ueta and Hiroshi Kanzaki and Koichi Kobayashi and Yutaka Toyozawa and Eiichi Hanamura},
title = {Excitonic Processes in Solids},
publisher = {Springer Berlin, Heidelberg},
year = {2012},
doi = {https://doi.org/10.1007/978-3-642-82602-3},
}

@book{YuCardona2010book,
  title     = {Fundamentals of Semiconductors: Physics and Materials Properties},
  author    = {Yu, Peter Y. and Cardona, Manuel},
  edition   = {4},
  year      = {2010},
  publisher = {Springer},
  doi       = {10.1007/978-3-642-00710-1}
}

@book{Agranovich2008book,
    author = {Agranovich, Vladimir},
    title = {Excitations in Organic Solids},
    publisher = {Oxford University Press},
    year = {2008},
    month = {11},
    isbn = {9780199234417},
    doi = {10.1093/acprof:oso/9780199234417.001.0001},
    url = {https://doi.org/10.1093/acprof:oso/9780199234417.001.0001},
}

@book{BornHuang1954book,
  title     = {Dynamical Theory of Crystal Lattices},
  author    = {Born, Max and Huang, Kun},
  year      = {1954},
  publisher = {Oxford University Press},
  address   = {Oxford, UK}
}

@article{Ritchie1957PR,
  title = {Plasma Losses by Fast Electrons in Thin Films},
  author = {Ritchie, R. H.},
  journal = {Phys. Rev.},
  volume = {106},
  issue = {5},
  pages = {874--881},
  numpages = {0},
  year = {1957},
  month = {Jun},
  publisher = {American Physical Society},
  doi = {10.1103/PhysRev.106.874},
  url = {https://link.aps.org/doi/10.1103/PhysRev.106.874}
}

@article{Batson1983PRB,
  title = {Experimental energy-loss function, $\mathrm{Im}[\ensuremath{-}\frac{1}{\ensuremath{\epsilon}}(q,\ensuremath{\omega})]$, for aluminum},
  author = {Batson, P. E. and Silcox, J.},
  journal = {Phys. Rev. B},
  volume = {27},
  issue = {9},
  pages = {5224--5239},
  numpages = {0},
  year = {1983},
  month = {May},
  publisher = {American Physical Society},
  doi = {10.1103/PhysRevB.27.5224},
  url = {https://link.aps.org/doi/10.1103/PhysRevB.27.5224}
}

@article{Nozieres1958PR,
  title = {Correlation Energy of a Free Electron Gas},
  author = {Nozi\`eres, P. and Pines, D.},
  journal = {Phys. Rev.},
  volume = {111},
  issue = {2},
  pages = {442--454},
  numpages = {0},
  year = {1958},
  month = {Jul},
  publisher = {American Physical Society},
  doi = {10.1103/PhysRev.111.442},
  url = {https://link.aps.org/doi/10.1103/PhysRev.111.442}
}

@article{Hopfield1958PR,
  title = {Theory of the Contribution of Excitons to the Complex Dielectric Constant of Crystals},
  author = {Hopfield, J. J.},
  journal = {Phys. Rev.},
  volume = {112},
  issue = {5},
  pages = {1555--1567},
  numpages = {0},
  year = {1958},
  month = {Dec},
  publisher = {American Physical Society},
  doi = {10.1103/PhysRev.112.1555},
  url = {https://link.aps.org/doi/10.1103/PhysRev.112.1555}
}

@article{Wannier1937PR,
  title = {The Structure of Electronic Excitation Levels in Insulating Crystals},
  author = {Wannier, Gregory H.},
  journal = {Phys. Rev.},
  volume = {52},
  issue = {3},
  pages = {191--197},
  numpages = {0},
  year = {1937},
  month = {Aug},
  publisher = {American Physical Society},
  doi = {10.1103/PhysRev.52.191},
  url = {https://link.aps.org/doi/10.1103/PhysRev.52.191}
}

@article{Rohlfing2000PRB,
  title = {Electron-hole excitations and optical spectra from first principles},
  author = {Rohlfing, Michael and Louie, Steven G.},
  journal = {Phys. Rev. B},
  volume = {62},
  issue = {8},
  pages = {4927--4944},
  numpages = {0},
  year = {2000},
  month = {Aug},
  publisher = {American Physical Society},
  doi = {10.1103/PhysRevB.62.4927},
  url = {https://link.aps.org/doi/10.1103/PhysRevB.62.4927}
}

@article{Baroni2001RMP,
  title = {Phonons and related crystal properties from density-functional perturbation theory},
  author = {Baroni, Stefano and de Gironcoli, Stefano and Dal Corso, Andrea and Giannozzi, Paolo},
  journal = {Rev. Mod. Phys.},
  volume = {73},
  issue = {2},
  pages = {515--562},
  numpages = {0},
  year = {2001},
  month = {Jul},
  publisher = {American Physical Society},
  doi = {10.1103/RevModPhys.73.515},
  url = {https://link.aps.org/doi/10.1103/RevModPhys.73.515}
}

@article{Wang2018RMP,
  title = {Colloquium: Excitons in atomically thin transition metal dichalcogenides},
  author = {Wang, Gang and Chernikov, Alexey and Glazov, Mikhail M. and Heinz, Tony F. and Marie, Xavier and Amand, Thierry and Urbaszek, Bernhard},
  journal = {Rev. Mod. Phys.},
  volume = {90},
  issue = {2},
  pages = {021001},
  numpages = {25},
  year = {2018},
  month = {Apr},
  publisher = {American Physical Society},
  doi = {10.1103/RevModPhys.90.021001},
  url = {https://link.aps.org/doi/10.1103/RevModPhys.90.021001}
}

@article{Gatti-Sottile2013PRB,
  title = {Exciton dispersion from first principles},
  author = {Gatti, Matteo and Sottile, Francesco},
  journal = {Phys. Rev. B},
  volume = {88},
  issue = {15},
  pages = {155113},
  numpages = {7},
  year = {2013},
  month = {Oct},
  publisher = {American Physical Society},
  doi = {10.1103/PhysRevB.88.155113},
  url = {https://link.aps.org/doi/10.1103/PhysRevB.88.155113}
}

@article{Onida2002RMP,
  title={Electronic excitations: density-functional versus many-body Green’s-function approaches},
  author={Onida, Giovanni and Reining, Lucia and Rubio, Angel},
  journal={Rev. Mod. Phys.},
  volume={74},
  number={2},
  pages={601},
  year={2002},
  doi={https://doi.org/10.1103/RevModPhys.74.601}
}

@article{Schulke1988PRB,
  title = {Interband transitions and core excitation in highly oriented pyrolytic graphite studied by inelastic synchrotron x-ray scattering: Band-structure information},
  author = {Sch\"ulke, W. and Bonse, U. and Nagasawa, H. and Kaprolat, A. and Berthold, A.},
  journal = {Phys. Rev. B},
  volume = {38},
  issue = {3},
  pages = {2112--2123},
  numpages = {0},
  year = {1988},
  doi = {10.1103/PhysRevB.38.2112},
}

@article{Schulke1989PRB,
  title = {Dynamic structure of electrons in Be metal by inelastic x-ray scattering spectroscopy},
  author = {Sch\"ulke, W. and Nagasawa, H. and Mourikis, S. and Kaprolat, A.},
  journal = {Phys. Rev. B},
  volume = {40},
  issue = {18},
  pages = {12215--12228},
  numpages = {0},
  year = {1989},
  doi = {10.1103/PhysRevB.40.12215},
}

@article{Watanabe2006BCSJ,
    author = {Watanabe, Noboru and Hayashi, Hisashi and Udagawa, Yasuo},
    title = "{Bethe Surface of Liquid Water Determined by Inelastic X-Ray Scattering Spectroscopy and Electron Correlation Effects}",
    journal = {Bulletin of the Chemical Society of Japan},
    volume = {70},
    number = {4},
    pages = {719-726},
    year = {2006},
    month = {06},
    issn = {0009-2673},
    doi = {10.1246/bcsj.70.719},
}

@article{Schulke1995PRB,
  title = {Dynamic and static structure factor of electrons in Si: Inelastic x-ray scattering results},
  author = {Sch\"ulke, W. and Schmitz, J. R. and Schulte-Schrepping, H. and Kaprolat, A.},
  journal = {Phys. Rev. B},
  volume = {52},
  issue = {16},
  pages = {11721--11732},
  numpages = {0},
  year = {1995},
  doi = {10.1103/PhysRevB.52.11721},
}

@article{Ishii2013JPSJ,
author = {Ishii ,Kenji and Tohyama ,Takami and Mizuki ,Jun'ichiro},
title = {Inelastic X-ray Scattering Studies of Electronic Excitations},
journal = {Journal of the Physical Society of Japan},
volume = {82},
number = {2},
pages = {021015},
year = {2013},
doi = {10.7566/JPSJ.82.021015},
URL = https://doi.org/10.7566/JPSJ.82.021015
}

@article{Weissker2010PRB,
  title = {Dynamic structure factor and dielectric function of silicon for finite momentum transfer: Inelastic x-ray scattering experiments and ab initio calculations},
  author = {Weissker, Hans-Christian and Serrano, Jorge and Huotari, Simo and Luppi, Eleonora and Cazzaniga, Marco and Bruneval, Fabien and Sottile, Francesco and Monaco, Giulio and Olevano, Valerio and Reining, Lucia},
  journal = {Phys. Rev. B},
  volume = {81},
  issue = {8},
  pages = {085104},
  numpages = {15},
  year = {2010},
  month = {Feb},
  doi = {10.1103/PhysRevB.81.085104},
}

@article{Chernikov2014PRL,
  title = {Exciton Binding Energy and Nonhydrogenic Rydberg Series in Monolayer ${\mathrm{WS}}_{2}$},
  author = {Chernikov, Alexey and Berkelbach, Timothy C. and Hill, Heather M. and Rigosi, Albert and Li, Yilei and Aslan, Burak and Reichman, David R. and Hybertsen, Mark S. and Heinz, Tony F.},
  journal = {Phys. Rev. Lett.},
  volume = {113},
  issue = {7},
  pages = {076802},
  numpages = {5},
  year = {2014},
  month = {Aug},
  publisher = {American Physical Society},
  doi = {10.1103/PhysRevLett.113.076802},
  url = {https://link.aps.org/doi/10.1103/PhysRevLett.113.076802}
}

@article{Alkauskas2013PRB,
  title = {Dynamic structure factors of Cu, Ag, and Au: Comparative study from first principles},
  author = {Alkauskas, Audrius and Schneider, Simon D. and H\'ebert, C\'ecile and Sagmeister, Stephan and Draxl, Claudia},
  journal = {Phys. Rev. B},
  volume = {88},
  issue = {19},
  pages = {195124},
  numpages = {11},
  year = {2013},
  month = {Nov},
  publisher = {American Physical Society},
  doi = {10.1103/PhysRevB.88.195124},
  url = {https://link.aps.org/doi/10.1103/PhysRevB.88.195124}
}

@article{Fugallo2015PRB,
  title = {Exciton energy-momentum map of hexagonal boron nitride},
  author = {Fugallo, Giorgia and Aramini, Matteo and Koskelo, Jaakko and Watanabe, Kenji and Taniguchi, Takashi and Hakala, Mikko and Huotari, Simo and Gatti, Matteo and Sottile, Francesco},
  journal = {Phys. Rev. B},
  volume = {92},
  issue = {16},
  pages = {165122},
  numpages = {5},
  year = {2015},
  month = {Oct},
  publisher = {American Physical Society},
  doi = {10.1103/PhysRevB.92.165122},
  url = {https://link.aps.org/doi/10.1103/PhysRevB.92.165122}
}

@article{Dagotto2005Science,
author = {Elbio Dagotto },
title = {Complexity in Strongly Correlated Electronic Systems},
journal = {Science},
volume = {309},
number = {5732},
pages = {257-262},
year = {2005},
doi = {10.1126/science.1107559},
URL = {https://www.science.org/doi/abs/10.1126/science.1107559},
}

@Article{Rossi2025xray,
author={Rossi, Thomas C.
and Qiao, Lu
and Dykstra, Conner P.
and Rodrigues Pela, Ronaldo
and Gnewkow, Richard
and Wallick, Rachel F.
and Burke, John H.
and Nicholas, Erin
and March, Anne Marie
and Doumy, Gilles
and Buchholz, D. Bruce
and Deparis, Christiane
and Z{\'u}{\~{n}}iga-P{\'e}rez, Jes{\'u}s
and Weise, Michael
and Ellmer, Klaus
and Fondell, Mattis
and Draxl, Claudia
and van der Veen, Renske M.},
title={Dynamic control of X-ray core-exciton resonances by Coulomb screening in photoexcited semiconductors},
journal={Communications Materials},
year={2025},
month={Aug},
day={20},
volume={6},
number={1},
pages={191},
issn={2662-4443},
doi={10.1038/s43246-025-00909-w},
url={https://doi.org/10.1038/s43246-025-00909-w}
}

@article{Leon2021PRB,
  title = {Frequency dependence in $GW$ made simple using a multipole approximation},
  author = {Leon, Dario A. and Cardoso, Claudia and Chiarotti, Tommaso and Varsano, Daniele and Molinari, Elisa and Ferretti, Andrea},
  journal = {Phys. Rev. B},
  volume = {104},
  issue = {11},
  pages = {115157},
  numpages = {16},
  year = {2021},
  doi = {https://doi.org/10.1103/PhysRevB.104.115157},
}

@article{Leon2023PRB,
  title = {Efficient full frequency GW for metals using a multipole approach for the dielectric screening},
  author = {Leon, Dario A. and Ferretti, Andrea and Varsano, Daniele and Molinari, Elisa and Cardoso, Claudia},
  journal = {Phys. Rev. B},
  volume = {107},
  issue = {15},
  pages = {155130},
  numpages = {15},
  year = {2023},
  month = {Apr},
  publisher = {American Physical Society},
  doi = {10.1103/PhysRevB.107.155130},
  url = {https://link.aps.org/doi/10.1103/PhysRevB.107.155130}
}

@article{Leon2025PRB,
  title = {Spectral properties from an efficient analytical representation of the $GW$ self-energy within a multipole approximation},
  author = {Leon, Dario A. and Berland, Kristian and Cardoso, Claudia},
  journal = {Phys. Rev. B},
  volume = {111},
  issue = {19},
  pages = {195147},
  numpages = {16},
  year = {2025},
  month = {May},
  publisher = {American Physical Society},
  doi = {10.1103/PhysRevB.111.195147},
  url = {https://link.aps.org/doi/10.1103/PhysRevB.111.195147}
}

@article{Marsili2021PRB,
  title = {Spinorial formulation of the $GW$-BSE equations and spin properties of excitons in two-dimensional transition metal dichalcogenides},
  author = {Marsili, Margherita and Molina-S\'anchez, Alejandro and Palummo, Maurizia and Sangalli, Davide and Marini, Andrea},
  journal = {Phys. Rev. B},
  volume = {103},
  issue = {15},
  pages = {155152},
  numpages = {20},
  year = {2021},
  month = {Apr},
  publisher = {American Physical Society},
  doi = {10.1103/PhysRevB.103.155152},
  url = {https://link.aps.org/doi/10.1103/PhysRevB.103.155152}
}

@article{Adamska2021PRB,
  title = {Bethe-Salpeter equation approach with electron-phonon coupling for exciton binding energies},
  author = {Adamska, Lyudmyla and Umari, Paolo},
  journal = {Phys. Rev. B},
  volume = {103},
  issue = {7},
  pages = {075201},
  numpages = {10},
  year = {2021},
  month = {Feb},
  publisher = {American Physical Society},
  doi = {10.1103/PhysRevB.103.075201},
  url = {https://link.aps.org/doi/10.1103/PhysRevB.103.075201}
}

@article{Giustino2017RMP,
  title = {Electron-phonon interactions from first principles},
  author = {Giustino, Feliciano},
  journal = {Rev. Mod. Phys.},
  volume = {89},
  issue = {1},
  pages = {015003},
  numpages = {63},
  year = {2017},
  month = {Feb},
  publisher = {American Physical Society},
  doi = {10.1103/RevModPhys.89.015003},
  url = {https://link.aps.org/doi/10.1103/RevModPhys.89.015003}
}

@article{Devreese2009RPP,
doi = {10.1088/0034-4885/72/6/066501},
url = {https://doi.org/10.1088/0034-4885/72/6/066501},
year = {2009},
month = {may},
publisher = {},
volume = {72},
number = {6},
pages = {066501},
author = {Devreese, Jozef T and Alexandrov, Alexandre S},
title = {Fröhlich polaron and bipolaron: recent developments},
journal = {Reports on Progress in Physics},
}

@article{Carbotte1990RMP,
  title = {Properties of boson-exchange superconductors},
  author = {Carbotte, J. P.},
  journal = {Rev. Mod. Phys.},
  volume = {62},
  issue = {4},
  pages = {1027--1157},
  numpages = {0},
  year = {1990},
  month = {Oct},
  publisher = {American Physical Society},
  doi = {10.1103/RevModPhys.62.1027},
  url = {https://link.aps.org/doi/10.1103/RevModPhys.62.1027}
}

@article{Gonze1997PRB,
  title = {Dynamical matrices, Born effective charges, dielectric permittivity tensors, and interatomic force constants from density-functional perturbation theory},
  author = {Gonze, Xavier and Lee, Changyol},
  journal = {Phys. Rev. B},
  volume = {55},
  issue = {16},
  pages = {10355--10368},
  numpages = {0},
  year = {1997},
  month = {Apr},
  publisher = {American Physical Society},
  doi = {10.1103/PhysRevB.55.10355},
  url = {https://link.aps.org/doi/10.1103/PhysRevB.55.10355}
}

@article{Cochran1959PRL,
  title = {Crystal Stability and the Theory of Ferroelectricity},
  author = {Cochran, W.},
  journal = {Phys. Rev. Lett.},
  volume = {3},
  issue = {9},
  pages = {412--414},
  numpages = {0},
  year = {1959},
  month = {Nov},
  publisher = {American Physical Society},
  doi = {10.1103/PhysRevLett.3.412},
  url = {https://link.aps.org/doi/10.1103/PhysRevLett.3.412}
}

@article{Caldwell2015Nanophotonics,
title = {Low-loss, infrared and terahertz nanophotonics using surface phonon polaritons},
author = {Joshua D. Caldwell and Lucas Lindsay and Vincenzo Giannini and Igor Vurgaftman and Thomas L. Reinecke and Stefan A. Maier and Orest J. Glembocki},
pages = {44--68},
volume = {4},
number = {1},
journal = {Nanophotonics},
doi = {doi:10.1515/nanoph-2014-0003},
year = {2015},
}

@article{Aryasetiawan1998RPP,
   author = {F. Aryasetiawan and O. Gunnarsson},
   title = {The {GW} method},
   journal = {Rep. Prog. Phys.},
   volume = {61},
   number = {},
   pages = {237},
   year = {1998},
   doi = {https://doi.org/10.1088/0034-4885/61/3/002}
}

@article{Reining2018wcms,
author = {Reining, Lucia},
title = {The GW approximation: content, successes and limitations},
journal = {WIREs Computational Molecular Science},
volume = {8},
number = {3},
pages = {e1344},
doi = {https://doi.org/10.1002/wcms.1344},
year = {2018}
}

@article{Bradlyn2024PRB,
  title = {Spectral density and sum rules for second-order response functions},
  author = {Bradlyn, Barry and Abbamonte, Peter},
  journal = {Phys. Rev. B},
  volume = {110},
  issue = {24},
  pages = {245132},
  numpages = {15},
  year = {2024},
  month = {Dec},
  publisher = {American Physical Society},
  doi = {10.1103/PhysRevB.110.245132},
  url = {https://link.aps.org/doi/10.1103/PhysRevB.110.245132}
}

@incollection{Fink1989AEEP,
title = {Recent Developments in Energy-Loss Spectroscopy**This article has been accepted as Habilitation Thesis by the Fakultät für Physik, Universität Karlsruhe},
editor = {Peter W. Hawkes},
series = {Advances in Electronics and Electron Physics},
publisher = {Academic Press},
volume = {75},
pages = {121-232},
year = {1989},
issn = {0065-2539},
doi = {https://doi.org/10.1016/S0065-2539(08)60947-6},
url = {https://www.sciencedirect.com/science/article/pii/S0065253908609476},
author = {Jörg Fink},
}

@article{Mitrano2024PRX,
  title = {Exploring Quantum Materials with Resonant Inelastic X-Ray Scattering},
  author = {Mitrano, M. and Johnston, S. and Kim, Young-June and Dean, M. P. M.},
  journal = {Phys. Rev. X},
  volume = {14},
  issue = {4},
  pages = {040501},
  numpages = {32},
  year = {2024},
  month = {Dec},
  publisher = {American Physical Society},
  doi = {10.1103/PhysRevX.14.040501},
  url = {https://link.aps.org/doi/10.1103/PhysRevX.14.040501}
}

@article{Ciraci2012Science,
author = {C. Ciracì  and R. T. Hill  and J. J. Mock  and Y. Urzhumov  and A. I. Fernández-Domínguez  and S. A. Maier  and J. B. Pendry  and A. Chilkoti  and D. R. Smith },
title = {Probing the Ultimate Limits of Plasmonic Enhancement},
journal = {Science},
volume = {337},
number = {6098},
pages = {1072-1074},
year = {2012},
doi = {10.1126/science.1224823},
URL = {https://www.science.org/doi/abs/10.1126/science.1224823},
}

@ARTICLE{Engel1993PRB,
   author = {G.~E. Engel and Behnam Farid},
   title = {Generalized plasmon-pole model and plasmon band structures of crystals},
   journal = {Phys. Rev. B},
   volume = {47},
   number={23},
   pages = {15931},
   year = {1993},
   doi = {https://doi.org/10.1103/PhysRevB.47.15931}
}

@ARTICLE{Hybertsen1986PRB,
   author = {Mark~S. Hybertsen and Steven~G. Louie},
   title = {Electron correlation in semiconductors and insulators: Band gaps and quasiparticle energies},
   journal = {Phys. Rev. B},
   volume = {34},
   number = {},
   pages = {5390},
   year = {1986},
   doi = {https://doi.org/10.1103/PhysRevB.34.5390}
}

@article{Zhang1989PRB,
  title = {Evaluation of quasiparticle energies for semiconductors without inversion symmetry},
  author = {Zhang, S. B. and Tom\'anek, D. and Cohen, Marvin L. and Louie, Steven G. and Hybertsen, Mark S.},
  journal = {Phys. Rev. B},
  volume = {40},
  issue = {5},
  pages = {3162--3168},
  numpages = {0},
  year = {1989},
  doi = {https://doi.org/10.1103/PhysRevB.40.3162},
}

@article{vonderLinden1988PRB,
  title = {Precise quasiparticle energies and Hartree-Fock bands of semiconductors and insulators},
  author = {{von der Linden}, Wolfgang and Horsch, Peter},
  journal = {Phys. Rev. B},
  volume = {37},
  issue = {14},
  pages = {8351--8362},
  numpages = {0},
  year = {1988},
  doi = {https://doi.org/10.1103/PhysRevB.37.8351},
}

@ARTICLE{Godby1989PRL,
   author = {R.~W. Godby and R.~J. Needs},
   title = {Metal-insulator transition in Kohn-Sham theory and quasiparticle theory},
   journal = {Phys. Rev. Lett.},
   volume = {62},
   number = {},
   pages = {1169},
   year = {1989},
   doi = {https://doi.org/10.1103/PhysRevLett.62.1169}
}

@Article{Vidberg1977JLTP,
author={Vidberg, H. J.
and Serene, J. W.},
title={Solving the Eliashberg equations by means ofN-point Pad{\'e} approximants},
journal={Journal of Low Temperature Physics},
year={1977},
month={Nov},
day={01},
volume={29},
number={3},
pages={179-192},
issn={1573-7357},
doi={10.1007/BF00655090},
url={https://doi.org/10.1007/BF00655090}
}

@article{Vos2017JPCS,
title = {Simple model dielectric functions for insulators},
journal = {Journal of Physics and Chemistry of Solids},
volume = {104},
pages = {192-197},
year = {2017},
issn = {0022-3697},
doi = {https://doi.org/10.1016/j.jpcs.2016.12.015},
url = {https://www.sciencedirect.com/science/article/pii/S0022369716311945},
author = {Maarten Vos and Pedro L. Grande},
}

@Article{Rivera2015NatComm,
author={Rivera, Pasqual
and Schaibley, John R.
and Jones, Aaron M.
and Ross, Jason S.
and Wu, Sanfeng
and Aivazian, Grant
and Klement, Philip
and Seyler, Kyle
and Clark, Genevieve
and Ghimire, Nirmal J.
and Yan, Jiaqiang
and Mandrus, D. G.
and Yao, Wang
and Xu, Xiaodong},
title={Observation of long-lived interlayer excitons in monolayer MoSe2--WSe2 heterostructures},
journal={Nature Communications},
year={2015},
month={Feb},
day={24},
volume={6},
number={1},
pages={6242},
issn={2041-1723},
doi={10.1038/ncomms7242},
url={https://doi.org/10.1038/ncomms7242}
}

@Article{Fogler2014NatComm,
author={Fogler, M. M.
and Butov, L. V.
and Novoselov, K. S.},
title={High-temperature superfluidity with indirect excitons in van der Waals heterostructures},
journal={Nature Communications},
year={2014},
month={Jul},
day={28},
volume={5},
number={1},
pages={4555},
issn={2041-1723},
doi={10.1038/ncomms5555},
url={https://doi.org/10.1038/ncomms5555}
}

@article{Rangel2016PRB,
  title = {Structural and excited-state properties of oligoacene crystals from first principles},
  author = {Rangel, Tonatiuh and Berland, Kristian and Sharifzadeh, Sahar and Brown-Altvater, Florian and Lee, Kyuho and Hyldgaard, Per and Kronik, Leeor and Neaton, Jeffrey B.},
  journal = {Phys. Rev. B},
  volume = {93},
  issue = {11},
  pages = {115206},
  numpages = {16},
  year = {2016},
  month = {Mar},
  publisher = {American Physical Society},
  doi = {10.1103/PhysRevB.93.115206},
  url = {https://link.aps.org/doi/10.1103/PhysRevB.93.115206}
}

@article{Li2025Vacuum,
  title = {First-Principles Calculations and Theoretical Analysis of {$\pi$} Plasmon in Graphene and Graphite: {{From 2D}} to {{3D}}},
  author = {Li, Pengfei and Hui, Ningju},
  year = {2025},
  journal = {Vacuum},
  volume = {240},
  pages = {114424},
  issn = {0042-207X},
  doi = {10.1016/j.vacuum.2025.114424},
}

\end{document}